# Single Ultrabright Fluorescent Silica Nanoparticles Can Be Used as Individual Fast Real-Time Nanothermometers

Mahshid Iraniparast, [1], Nishant Kumar, [1] Igor Sokolov [1,2,3,*]

[1] Department of Mechanical Engineering, Tufts University, Medford, MA 02155, USA;

[2] Department of Biomedical Engineering, Tufts University, Medford, MA 02155, USA.

[3] Department of Physics, Tufts University, Tufts University, Medford, MA 02155, USA.

ABSTRACT: Optical-based nanothermometry represents a transformative approach for precise temperature measurements at the nanoscale, which finds versatile applications across biology, medicine, and electronics. The assembly of ratiometric fluorescent 40 nm nanoparticles designed to serve as individual nanothermometers is introduced here. These nanoparticles exhibit unprecedented sensitivity (11% /K) and temperature resolution (128 $K \cdot Hz^{-1/2} \cdot W \cdot cm^{-2}$), outperforming existing optical nanothermometers by factors of 2-6 and 455, respectively. The enhanced performance is attributed to the encapsulation of fluorescent molecules with high density inside the mesoporous matrix. It becomes possible after incorporating hydrophobic groups into the silica matrix, which effectively prevents water ingress and dye leaking. -A practical application of these nanothermometers is demonstrated using confocal microscopy, showcasing their ability to map temperature distributions accurately. This methodology is compatible with any fluorescent microscope capable of recording dual fluorescent channels in any transparent medium or on a sample surface. This work not only sets a new benchmark for optical nano-thermometry but also provides a relatively simple yet powerful tool for exploring thermal phenomena at the nanoscale across various scientific domains.

# Introduction

Nanothermometry, a burgeoning research area, is pivotal for fundamental and applied sciences, enabling unprecedented insights into temperature dynamics at the nanoscale. Measurement of temperature at the nanoscale [1-4] opens new horizons in various fields of study, ranging from biological science [5-7] to material science [8, 9] and electrical engineering [10, 11]. For example, the temperature distribution in the cellular medium provides us with important information about the complicated biological and biochemical processes taking place within each cells [12,13, 14,15]. The measurements of nanoscale hotspots in semiconductors as well as the thermal imaging of integrated circuit devices, demonstrate the necessity of nanoscale sensors in the engineering field [10, 11].

Fluorescence of materials can depend on temperature. The fluorescence of individual molecules is typically not high enough to be easily detectable, suffers from photobleaching, and may depend on the environment (which nanoscale structure is typically unknown). Encapsulation of fluorophores inside of nanoparticles is a path to addressing the above problems. There are different strategies to create nanoparticles, in which fluorescence can be a measure of temperature. The specific ways of correlating temperature and fluorescence include the emission intensity, the excitation-emission ratio, the ratio of the fluorescence intensities at different wavelengths, the position of the intensity maximum, and the fluorescence lifetime. There are also optical ways to measure temperature-dependent electron spin resonance and magnetic resonances of the particle material. Among the nanoparticles suitable for such measurements, there are particles with encapsulated small organic molecules and lanthanide complexes. The nanoparticles also include inorganic quantum dots, vacancy-containing nanodiamonds, carbon dots, etc., see [12] for detail. Ref. [12] has also a detailed analysis of the advantages and disadvantages of various particles-nanothermometers (nanothermometers).

Initially, suspensions of nanoparticles were investigated for temperature measurement [16-18]. Those particles were not sufficiently bright to resolve individual nanoparticles at a sufficient signal-to-noise ratio. As a result, the spatial resolution achieved in those works was at the level of multi-micron at best (the size of the volume that produces sufficient fluorescence intensity). The ability to measure the fluorescence of individual nanothermometers has the potential to achieve better spatial resolution up to the particle size. High fluorescence brightness also allows for a high temporal resolution [19-22] because it allows collecting reliably detectable number of photons (signal) faster.

There are several challenges related to the utilization of individual nanoparticles as nanothermometers [23, 24]. The problems of high fluorescence brightness, relative sensitivity, and spatiotemporal resolution have recently been identified as the major challenges in fluorescent ratiometric nanothermometry [12]. In addition, the limited reproducibility of individual nanoparticles is one of the challenges to their widespread applications [12, 25]. For example, individual nanoparticles require higher excitation intensity compared to particle suspensions, which may result in the self-heating of individual nanoparticles and the surroundings. This may lead to bias in temperature measurements and deterioration of the accuracy as well as the change in the measured system itself [19, 26]. These issues need to be addressed to effectively utilize individual nanoparticles as thermometers.

Here, we demonstrate a way of addressing the above issues. We introduce ultrabright fluorescent silica nanothermometers (UNSNTs) that can be used as individual nanothermometers with a record-high spatiotemporal resolution, with minimal spectral variations between individual nanothermometers. Note that the spatial and temporal resolutions are mutually dependent followed by the ergodic hypothesis of thermodynamics; as an example, the spatial resolution can be the size of individual UNSNTs, whereas a sufficient fluorescence signal-to-noise ratio is achieved during 0.07 msec integration time, see Figure 3 of this paper for detail. A more relevant characteristic describing the spatiotemporal resolution of individual nanothermometers is the temperature resolution. As we demonstrate here, the temperature resolution of our UNSNTs temperature resolution is $128 \text{ K} \cdot \text{Hz}^{-1/2} \cdot \text{W} \cdot \text{cm}^{-2}$. It outperforms the existing optical nanothermometers by a factor of 455. We further demonstrate that the temperature reading using UNSNTs is independent of the power of the excitation light and the particle volume. The latter is important; although UNSNTs are relatively monodispersed (with an average size of 40 nm), there is always a possibility of particle aggregation.

Individual UNSNTs have been chosen in this study because of their exceptional brightness, which does not require high excitation power. Furthermore, the assembly of these particles involves the encapsulation of fluorescent dyes without their chemical modification [27-30]. Finally, the fluorescent dyes show excellent photostability after the encapsulation [31]. Two different fluorescent dyes can be encapsulated within these nanoparticles, which makes them an excellent candidate for fluorescent ratiometric nanothermometers [32]. It has been shown that the fluorescence spectra of such particles can be designed with high accuracy [33]. Furthermore, spectral ratios of these fluorescent particles demonstrated rather little variations between different particles, which was shown for multiple dye ratios [33]. The fluorescence of nanothermometers presented here has been obtained by encapsulation of Rhodamine 6G (R6G) (temperature independent) and Rhodamine B (RB) (temperature dependent) dyes. To characterize the brightness of our particles, we use the units broadly used in flow cytometry to characterize the brightness of complex fluorophores. Specifically, we use MESF units (Molecules of Equivalent Soluble Fluorochrome), which are relative to standard fluorophores (see, the Supplementary Information, Calculating the fluorescence brightness of the nanothermometers section, for detail. It was found that the brightness of our 40 nm particles was equivalent to the fluorescence brightness of 650 molecules of R6G and 1020 of RB dyes. This is substantially brighter than the brightness of quantum dots of similar spectra.

Compared to the previously reported mesoporous silica nanothermometers with similar encapsulated dyes [31], UNSNTs contain a much higher density of the encapsulated dyes. The higher retention of the encapsulated dyes is attained by using the method reported in [34], incorporating hydrophobic groups in the silica matrix. It creates a strong hydrophobic interior in the mesoporous silica matrix [35] while maintaining a relatively hydrophilic surface of UNSNTs (the particles are well dispersed in the aqueous environment). This use of geometry to enhance hydrophobicity is similar to the known Lotus effect; it effectively prevents the leakage of fluorescent dye molecules and, therefore, results in higher retention of the encapsulated fluorescent dyes inside UNSNTs.

There are a number of reports of sensitive nanothermometers in the literature. It has been suggested that relative sensitivity and thermal resolution be used as the main characteristics of nanothermometers. For example, the relative sensitivity of 6.9% $K^{-1}$ was reported for CdSe/CdS$_x$Se$_{1-x}$ quantum dots [36], 2.49% $K^{-1}$ for CdSeS/ZnS in the matrix of poly(methyl methacrylate-co-methacrylic acid) [37], and 1.8% $K^{-1}$ for co-doped nanodiamonds based on GeV/SiV [38].

The important characteristic of nanothermometers is their thermal resolution. The relative sensitivity is the same, no matter if the temperature is measured by using one or a million thermometers together, using a high or low intensity of the excitation light, using milliseconds or minutes. Therefore, it is important to introduce a characteristic that would be specific to the thermal characteristics of individual nanothermometers. It is the thermal resolution. Since the thermal fluctuations in the nanoscale are substantial, one needs a sufficient "number of measurements" or to average out the thermal fluctuations to reach the desirable accuracy. The thermal resolutions are averaged out faster when the power density of the excitation light is higher (because more photons are participating in the measurements). To avoid the artificial improvement of the thermal resolution by just using a higher power density of the excitation light, we use the thermal resolution normalized by the excitation power as suggested in [38]. The excitation power densities used in this study were 425 (W.cm$^{-2}$), which are lower than those used in the literature before [38]. Despite the lower power densities employed, our results consistently yielded enhancement in the temperature resolution of individual nanothermometers compared to the previously reported results [19, 38, 39]. Specifically, the temperature resolution of UNSNTs is 128 $K \cdot Hz^{-1/2} \cdot W \cdot cm^{-2}$, representing more than 455x enhancement compared to the record reported before [38]. The sensitivity of UNSNTs is 11% $K^{-1}$, demonstrating 6x enhancement compared to the values reported for the previous record [38].

# Results and Discussion

The geometrical characteristics of UNSNTs are shown in Figure 1. The nanoporous silica matrix of UNSNTs was created using a templated self-assembly as described in [29, 40]. The size distribution of individual nanoparticles and overall geometry were investigated using both the Dynamic Light Scattering (DLS) method and Atomic Force Microscopy (AFM) imaging. AFM images of nanothermometers show a close-to-spherical geometry of nanoparticles (Figure 1a). Figure 1b presents the distribution of the particle size derived from the AFM image. One can see that the majority of nanoparticles are within the size range of 20 to 65 nm. This finding is well aligned with the DLS result, which indicates that 70% of nanoparticles fell within the same size range of 20 to 65 nm (Figure 1c). AFM images give additional information about the geometry of the particles, which is apparently spherical.

The optical properties of UNSNTs dispersed in water are presented in Figure 2. Figure 2a shows a schematic of the photonic structure of nanothermometers. Two different fluorescent dyes, R6G and RB, were encapsulated inside nanochannels of silica matrix. The fluorescence spectra of the water dispersion of nanothermometers for different temperatures are shown in Figure 2b. RB dye is a well-known temperature-dependent fluorescent dye. As expected, the part of the spectra that corresponds to RB dye shows the dependence on temperature. To measure the temperature using the ratiometric approach, we integrated the fluorescence intensities within the following spectral ranges: 510 nm - 560 nm (spectral range of R6G dye) and 560 nm - 580 nm (spectral range of RB dye). A linear dependence of the ratio of temperature was observed, Figure 2c.

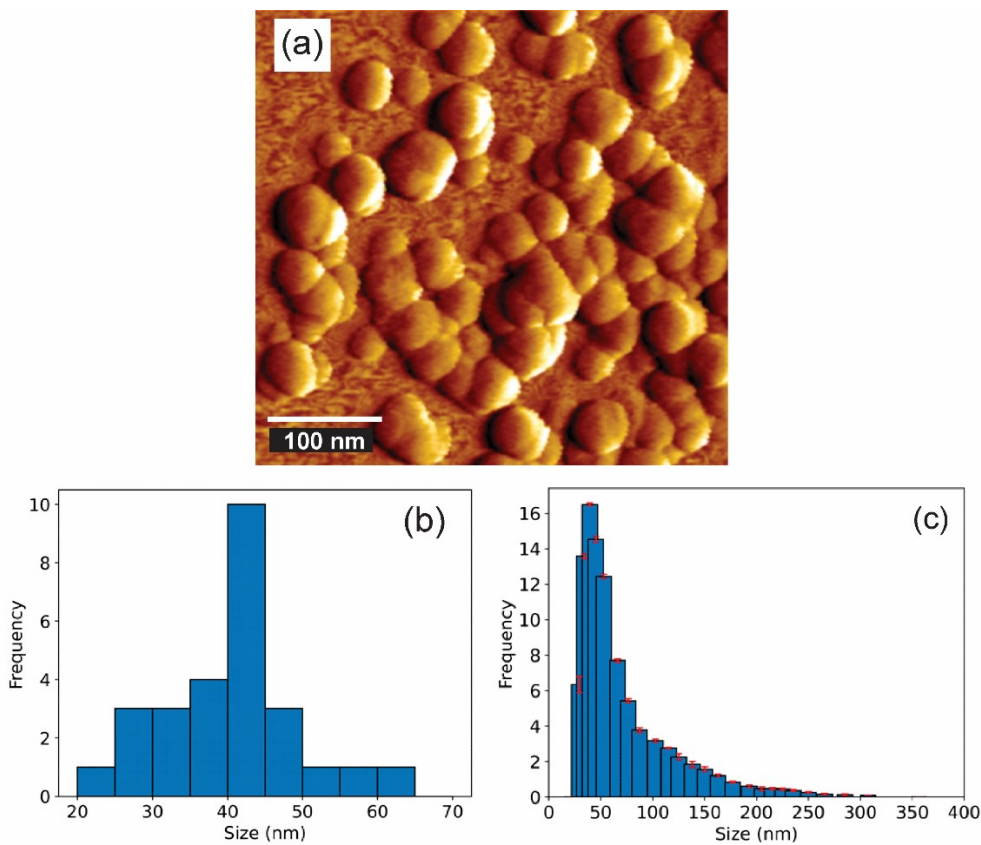

**Figure 1. The geometrical characteristics of nanothermometers (UNSNTs). (a) An AFM image of nanothermometers showing the nearly round morphology of UNSNTs. (b) The particle size distribution derived from the AFM images. (c) The particle size distribution obtained with the dynamic light scattering (DLS) technique.**

In these measurements, a single excitation wavelength of 488nm was used to excite mainly R6G fluorescent dye (temperature-independent dye), which transferred a significant portion of its energy to RB (temperature-dependent dye) via the Förster Resonance Energy Transfer (FRET) [41] because of close proximity of these two types of dye molecules. In this process, R6G was the donor, and RB was the acceptor of the FRET pair. The presence of FRET can be seen in multiple measurements, see, section "Demonstration of the Förster resonance energy transfer (FRET) between the dyes encapsulated inside nanothermometers" of the Supplementary information.

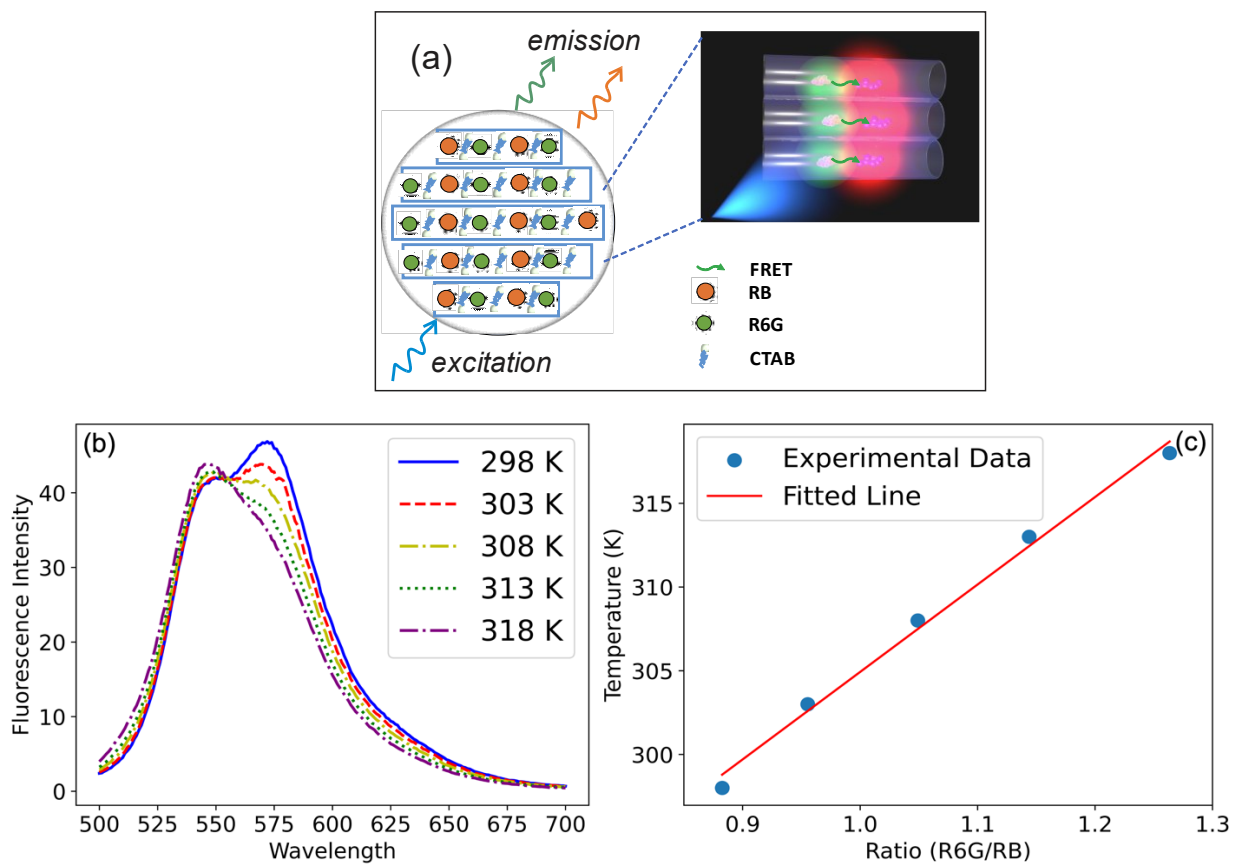

Figure 2: The optical properties of nanothermometers (UNSNTs). (a) A schematic of the structure of nanothermometers: nanoporous silica matrix with encapsulated two fluorescent dyes. 488nm light excites mainly R6G fluorescent dye (donor temperature independent dye), which transfers its energy via FRET to RB (donor temperature dependent dye). (b) Fluorescence spectra of the nanothermometers dispersed in water. The part of the spectra that corresponds to RB shows the dependence on temperature. (c) The ratio of fluorescence intensities versus temperature of water dispersion of nanothermometers averaged within 510 nm - 560 nm and 560 nm - 580 nm.

The ultrabright nature of the synthesized nanothermometers also allows measuring the fluorescence spectrum of individual UNSNTs. These measurements were done here using a Raman confocal microscopy setup, which allows to measure the entire fluorescence spectrum coming from individual nanoparticles (to avoid confusion, we did not measure the Raman signal here, but just used the Raman confocal microscopy setup to measure the entire fluorescence spectrum at each pixel of the image of individual UNSNTs). This setup is shown in Figure 3a. The results of characterizing the photonic properties of individual nanothermometers are presented in the other panels of Figure 3. The spectra of 10 individual nanothermometers excited at 488 nm (the acquisition time was 30 ms for each spectrum per nanoparticle) are shown in Figure 3b. One can see the uniformity of the spectra. This is key for the use of these nanoparticles as individual nanoparticle-based UNSNTs.

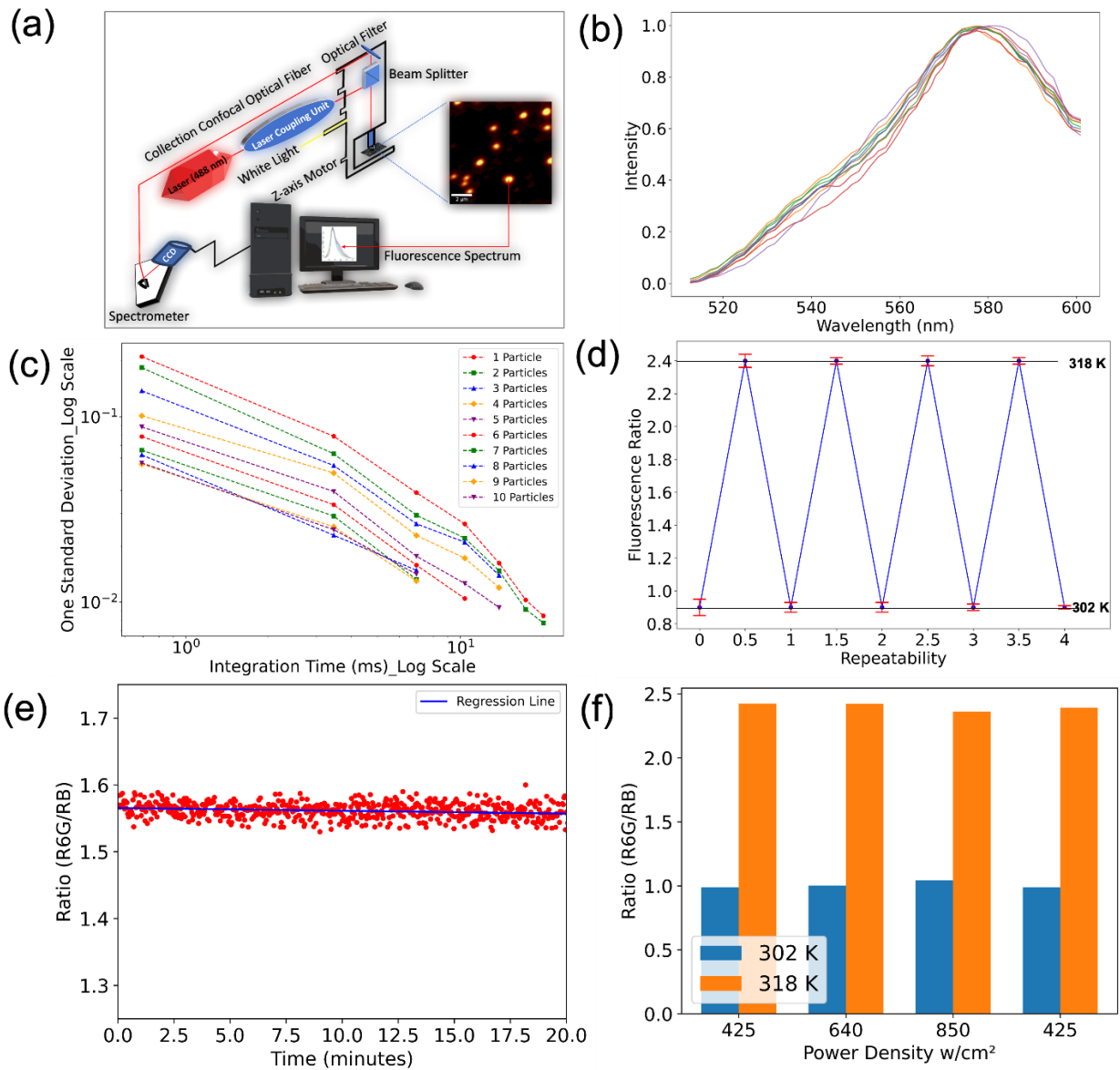

Figure 3: Photonic characteristics of individual nanothermometers (UNSNTs). (a) A schematic of the Raman confocal microscopy setup used to record fluorescence spectra of individual UNSNTs. (b) The spectral consistency of single UNSNTs. The fluorescence spectra of ten individual UNSNTs with an acquisition time of 30 ms are presented. The excitation wavelength is 488 nm. (c) Temperature fluctuations (the shot noise) around 302 K versus the integration time averaged for N UNSNTs as a function of integration time (N=1..10). (d) Stability of individual UNSNTs measured up to four full thermal cycles between 302 K and 318 K. The integration time of measurements was 20 ms per particle (was accumulated through 300 frames; 0.07 ms per frame) (e) A long duration of continuous measurements of the R6G/RB fluorescence ratio of an individual UNSNT at room temperature over 20 min. The integration time for each data point (UNSNT) is 0.07 ms. (f) The fluorescence ratio of R6G to RB averaged over five individual UNSNTs at different power densities at two different temperatures.

To find the temperature resolution and verify the absence of bias in the fluorescence measurements, it is instructive to vary the integration time for the fluorescence measurements. Provided no bias, the random noise should be averaged out so the standard deviation from the average is inversely proportional to the square root of the measurement time. Figure 3c shows a standard deviation of the measured 302 K temperature as a function of integration time in the log-log scale. In this scale, the power dependence should look like a straight line. One can see a rather good match (see the Supplementary information, Fig. S1a, for details). A small deviation from the power law at small measurement times can be explained by insufficiently large number of measured photons. Figure 3c shows the results which were averaged on N nanothermometers (N=1..10).

Another essential characteristic of nanoparticle-based thermometers is the repeatability of temperature measurements with respect to multiple cycles in the temperature change. The results of four full thermal cycles between 302 K and 318 K are shown in Figure 3d. Similar to the bulk measurements of the dependence of temperature on the fluorescence ratio (Figure 2c), the dependence for individual nanoparticles is also linear and can be approximated by the following equation: $Temperature = 10.3 \times Ratio + 293$, see Supplementary information Figure S2.

Another important characteristic, the consistency of R6G/RB fluorescence ratio over time, is shown in Figure 3e. This was confirmed by continuously measuring the fluorescence ratio for 20 minutes. The fluorescence photostability of each dye is shown in Supplementary information Figure S4. Each ratio measurement represents one individual nanoparticle whose fluorescence ratio was recorded for 0.07 msec at each shown point (for the entire particle; compare with 50 ms used for the previously reported record [38]). A small drift of the ratio over 20 minutes can be seen in Figure 3e. A linear regression was performed to find the change on the fluorescence ratio ratio, yielding the following equation: $ratio = -0.00040 time[min] + 1.57$. This results in the drift of the fluorescence ratio to 0.008 after twenty minutes of continuous scanning. According to the equation of temperature-ratio conversion, the change in temperature ($\delta T = 10.3 \times ratio$), the drift of the temperature was just 0.08 K. This drift is negligible, considering the observed temperature fluctuations. Therefore, it is safe to claim that the nanothermometers remain stable within at least 20 min of continuous measurements. It should be noted that 20 minutes of continuous measurements were recorded through 600 sequential images of the particles. This level of data collection should be sufficient for even the most detailed long-term studies, as excitation light is only switched on briefly during imaging.

Next, it is important to verify the independence of the fluorescence ratio of the intensity of the excitation light used. The higher power density of the excitation light can potentially lead to the heating of individual nanoparticles due to the absorbance of the excitation light. This may compromise the accuracy of temperature measurements. Secondly, the entire idea of ratiometric sensing is independence of the ratio of the intensity of the excitation light. To investigate the impact of power density on the fluorescence ratio of UNSNTs, three different intensities were used for two different temperatures (302 K and 318 K). Figure 3f shows that the excitation power density did not significantly affect the ratio of R6G to RB, indicating that self-heating was not an issue within the range of intensities used in the experiment. The same panel also shows the reversibility of the fluorescent ratio of R6G to RB to the same values after returning to the initial power density of 425 (W · cm$^{-2}$) at both high and low temperatures. It demonstrates the repeatability of the temperature measurements using UNSNT particles. (The laser power density was measured as described in the Supplementary information.)

As we discussed before, the most important characteristic of the thermometers is the temperature resolution $\eta_T$. It is defined as the numerator in the following equation [38]: $\sigma_T = \eta_T/\sqrt{t_m}$, where $\sigma_T$ represents the temperature fluctuations or uncertainty (calculated as one standard deviation from the average), $t_m$ is the integration time of collecting the fluorescence signal. Thus, the temperature resolution can be found by calculating the temperature uncertainty for different integration times. It can be found using the data example shown in Figure 3c. The calculation of the temperature resolution was performed for 60 individual nanoparticles to validate the results; see Supplementary information Figure S1. The average temperature resolution was 0.30 K·Hz$^{-1/2}$, with a mode of 0.25 K·Hz$^{-1/2}$.

When comparing these results with the reported in the literature, the power density of the excitation light must be considered because the temperature fluctuations $\sigma_T$ decreases when the power density of the excitation light increases. Following [38], we consider the temperature resolution multiplied by the power density. An excitation power density of 425 (W · cm$^{-2}$) was used for measuring temperature using individual UNSNTs. It results in a calculated temperature resolution multiplied by a power density of 128 K · Hz$^{-0.5}$ · W · cm$^{-2}$.

To benchmark our UNSNTs against the optical nanoparticle-based methods [38, 42-44], as suggested in [38], we plot the temperature resolution multiplied by power density versus the relative sensitivity, Figure 4b. The relative sensitivity (relative change of the fluorescence ratio per degree) is defined as $(\frac{\Delta R}{R})/\Delta T$, where R is the fluorescence ratio, $\Delta T$ is the change of temperature, and $\Delta R$ is the corresponding change of the ratio. The measured relative intensity of our individual nanothermometers was found to be $11 \pm 1.3 \ \%K^{-1}$, surpassing previous records in the literature. For example, 6.9% $K^{-1}$ was reported for CdSe/CdS$_x$Se$_{1-x}$ quantum dots [36], 2.49%$K^{-1}$ for CdSeS/ZnS in the matrix of poly(methyl methacrylate-co-methacrylic acid) [37], and 1.8% $K^{-1}$ for co-doped nanodiamonds based on GeV/SiV [38].

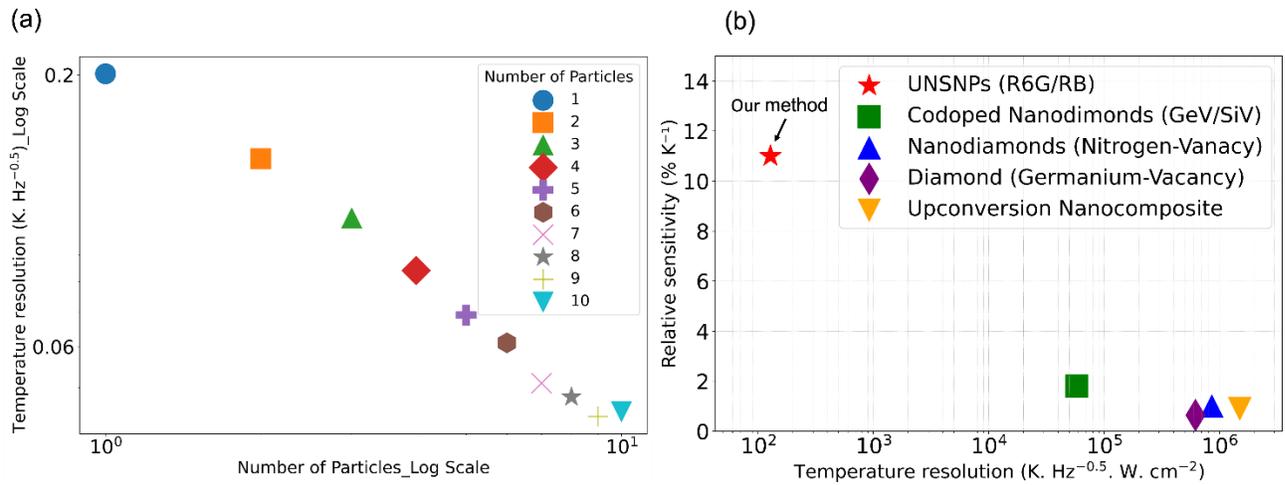

**Figure 4: Nanothermometers as sensitive and fast single-nanoparticle ratiometric nanothermometers (UNSNTs). (a) The temperature resolution as a function of the number of particles was calculated accordingly based on $\sigma_T = \eta_T/\sqrt{t_m}$, where $\sigma_T$ represents the temperature uncertainty, $t_m$ is the integration time, and $\eta_T$ is the temperature resolution. (b) Comparison to the previous art. Relative sensitivity versus temperature resolution multiplied by power density of the excitation light measured for UNSNTs, co-doped nanodiamonds based on GeV/SiV [38], nanodiamonds based on nitrogen-vacancy [42], diamond as a temperature sensor based on germanium vacancy [43], and carbon-coated core-shell up-conversion nanocomposite [44].**

As one can see from Figure 4b, the reported here nanothermometers show ~2-6x enhancement in the relative sensitivity and 455 times improvement in the temperature resolution multiplied by power density [38] compared to the previous record (based on both relative sensitivity and the temperature resolution (energy-related as explained above). This implies a new record.

It is instructional to confirm the homogeneous spread of the molecules directly through the particle volume. This is important for independence of the fluorescence spectrum on the volume of the particles. This analysis is presented in Figure 5. The Raman confocal microscopy setup (as shown in Figure 3a) was used to record the entire fluorescence spectrum at each pixel image (working, essentially, as a fluorescence microscope/spectrometer). It allowed to investigate a possible dependence of the fluorescence ratio of individual nanothermometers of different sizes/volumes. Figure 5a shows the fluorescence spectra collected from particles of different intensities. As an example, the intensity of the particles, shown in Figure 5a, varies up to five times. The same figure shows the fluorescence spectra collected from these particles (plotted in a logarithmic scale). One can see that the spectra remain virtually the same, just equally shifted with respect to each other (in a linear manner in the logarithmic scale). This demonstrates that the particles of different fluorescence intensities have the same spectrum multiplied by the number of molecular moieties encapsulated in each nanothermometer. This proves that the ratio between the numbers of the two encapsulated dyes and the average distance between dye molecules remain the same in the particles of different brightness, i.e., different volumes (otherwise, the spectra would be different due to the change of FRET efficiency). For the statistical proof of the direct proportionality between the volume and fluorescence brightness of the particles, we compare the distribution of the fluorescence brightness and the volumes of nanoparticles calculated from the DLS data (Figure 1c). The distributions of UNSNT volumes and fluorescent brightnesses are shown in Figure 5b. The similarity between these two plots demonstrates that the fluorescence intensity of UNSNTs is proportional to their volume.

Now, let us demonstrate a practical way of measuring temperature using individual UNSNTs with the help of laser scanning confocal microscopy, Figure 6a. Due to the presence of fluorescence resonance energy transfer (FRET) between the donor molecule (R6G) and the acceptor molecule (RB), the RB window exhibited a higher intensity compared to the R6G dye molecules. As a result, the detection window for R6G was deliberately set to be wider (510 nm to 560 nm) than that of RB molecules (560 nm to 580 nm). Figure 6b shows the fluorescence images of individual nanothermometers immobilized on a glass slide at the temperature of 308 K. The rightmost image is the ratio of the fluorescence intensities, which is proportional to temperature (see Supplementary information, Fig. S2). One can see a rather homogeneous distribution of the ratio (temperature) across the sample, which is expected.

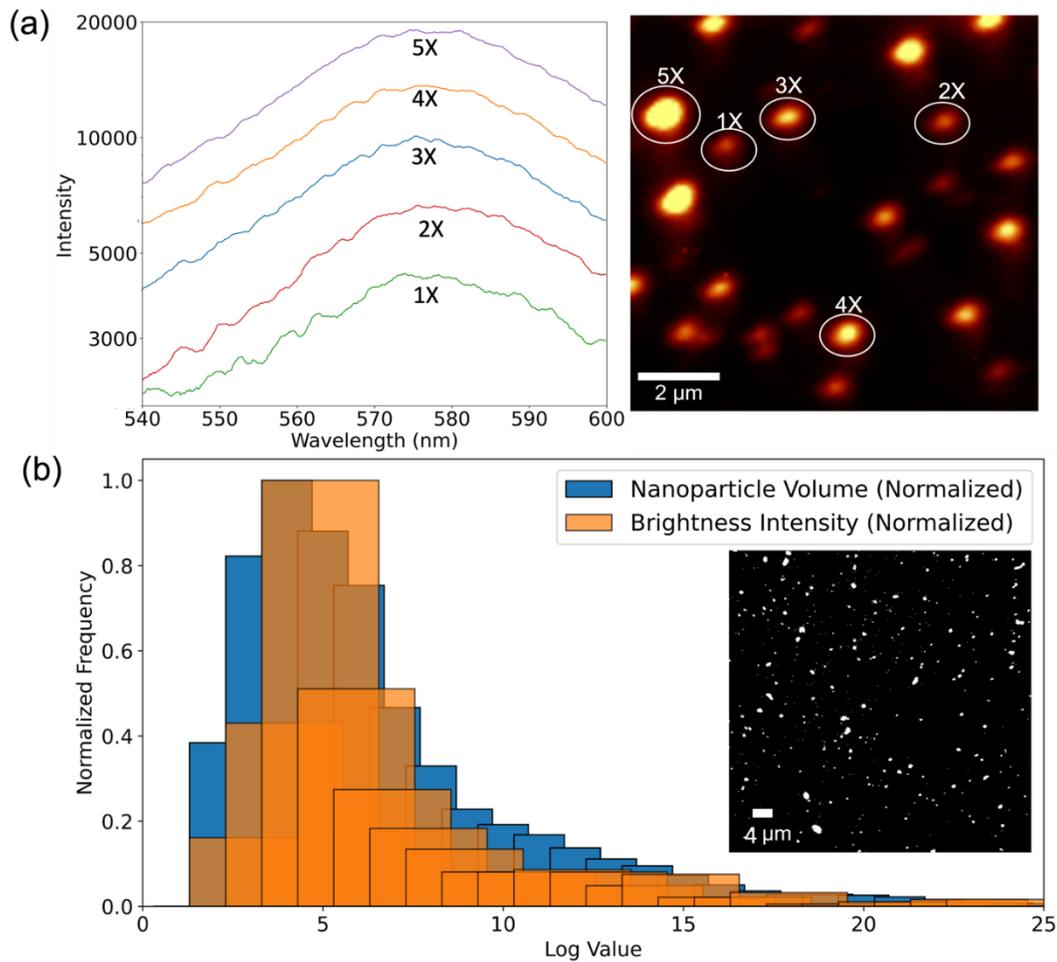

**Figure 5.** Demonstration of the even spread of the encapsulated dye molecules through the volume of UNSNTs. (a) Independence of the spectral shape on the UNSNT size. Fluorescent spectra of UNSNTs at different intensities showed the same spectral shape. The insert demonstrates the physical location of UNSNTs used to measure the spectra. The scalebar in the insert is 2 microns. (b) Distributions of the volume and fluorescence brightness of individual UNSNTs. The histograms of volume and intensity values of UNSNTs are shown. The similarity between these two plots demonstrates statistically that the fluorescence intensity of UNSNTs is proportional to their volume. The inserts show the location of the particles used for the calculation of fluorescence brightness.

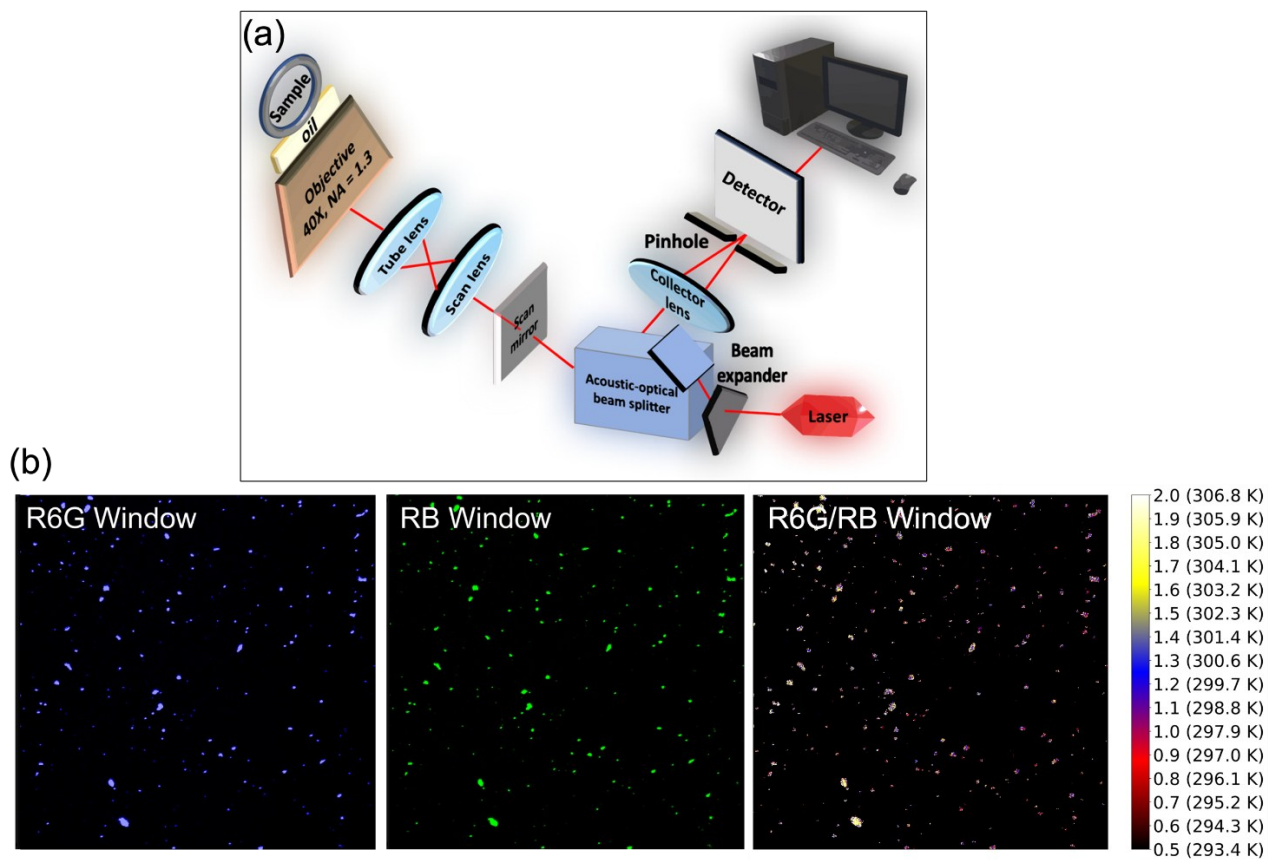

**Figure 6.** Measuring temperature with individual nanothermometers (UNSNTs) using laser scanning confocal microscopy. (a) A schematic of an inverted setup of the confocal microscopy used in the study to record images in two spectral windows. The excitation and detection channel ranges were determined based on the absorbance and fluorescence emission characteristics of the R6G and RB dye molecules: 500 nm excitation, 510 nm - 560 nm (corresponding to maximum fluorescence of R6G dye) and 560 nm - 580 nm (corresponding to maximum fluorescence of RB dye); the time of single acquisition per particle was 0.07 ms. (b) A 240 x 240 μm$^2$ image of individual UNSNTs (and their clusters). The images were collected through the R6G and RB fluorescence channels. The rightmost image is the ratio of fluorescent intensities converted to temperature. (The ratio of the background is artificially assigned to zero).

The same microscopy setup can be used to measure nontrivial distributions of temperature. It is shown in Figure 7. A coverslip glass was placed on a hot copper disc with a hole in the middle. This created a noticeable gradient of temperatures, as verified by the thermal imaging camera. The temperature was also measured using our nanothermometers deposited on the cover slip. The temperature was measured at three distinctive points across the sample: right in the center of the copper disc hole, at the edge of the coverslip adjacent to the copper disc, and on the copper disc, see Figure 7a. Figure 7b shows the temperature and fluorescence ratios defined by UNSNTs as well as the temperature measured by the thermal image camera. Based on these multiple measurements, one can estimate the difference between the temperature measurement within UNSNTs and the camera, which does not exceed 1 K, which is within the accuracy specifications of the thermal camera (absolute accuracy is ±2 K, and the relative accuracy is 0.07 K in the 253-573 K range).

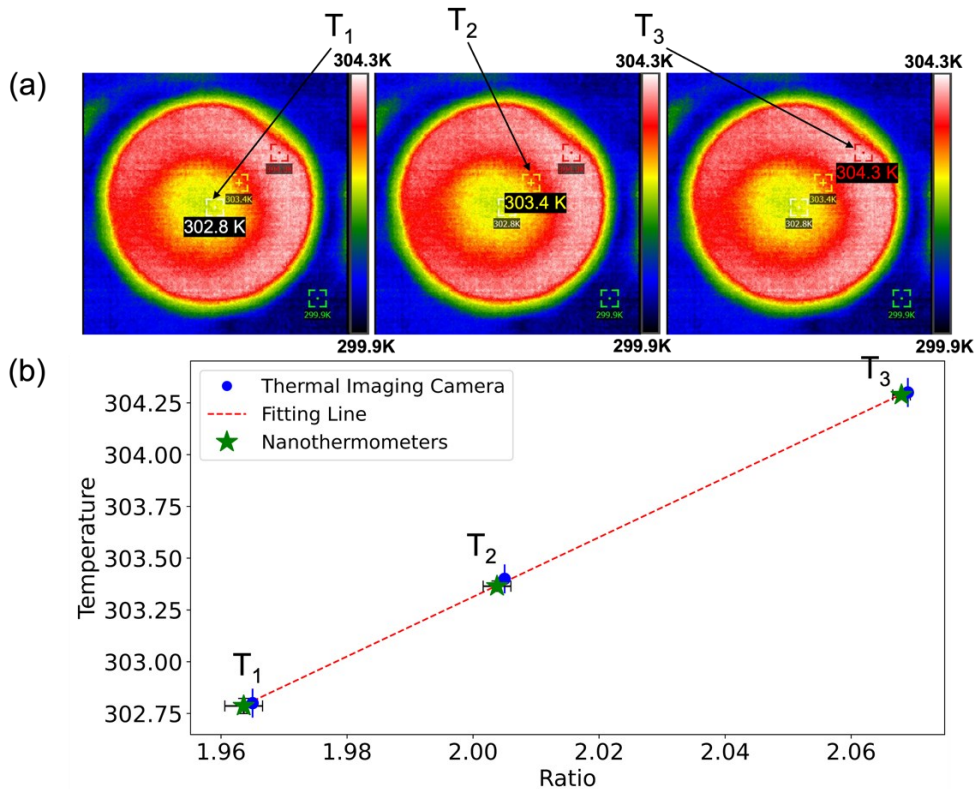

**Figure 7. The measurements of temperature distribution using a confocal laser scanning microscope. A cover slip glass was placed on a hot copper disc with a hole in the middle. The temperature gradient within a single frame, spanning from the edge to the center of the sample, was measured using both nanothermometers and thermal imaging camera at three different temperatures. (a) The thermal images and the location of the temperature measurements. Each image is 45 x 45 mm$^2$. (b) The plot of the measured temperatures versus the fluorescence ratio. It demonstrates the match between the temperature measurements obtained from nanothermometers (the average and one standard deviation are shown using green stars) and when using the thermal imaging camera (shown by blue-filled circles; the error bar indicates the relative accuracy of the thermal imager).**

It is important to note that nanoporous silica nanothermometers utilizing these two dyes were previously reported in [32]. However, the number of dye molecules encapsulated inside those particles was significantly lower. Here, we modified the synthesis by incorporating a small number of hydrophobic (ethyl) groups into the silica matrix. These additional groups create an "inverse Lotus effect" in the nanochannels, preventing water penetration and consequently preventing the densely packed encapsulated dye from leaking out into the aqueous environment [34]. As a result, the numbers of R6G and RB molecules encapsulated per ~40 nm - diameter UNSNT are now $1270 \pm 170$ and $1530 \pm 200$, respectively; see the Supplementary information for details. This is substantially more than what was reported before (360 and 330 of R6G and RB molecules, respectively, counted per 40 nm particle) [32]. When measuring the effective brightness of UNSNT, we found that a 40 nm diameter particle shows a fluorescence brightness equivalent to $650 \pm 20$ of R6G and $1020 \pm 30$ RB molecules.

Based on these results, we can speculate about the mechanism of the enhanced temperature resolution and sensitivity. The dependence of quantum yield of RB dye on temperature is well-known [32]. However, the observed temperature sensitivity reported in [32] was only 1%, which is more than ten times lower compared to the reported here. The main difference between the particles reported in [32] and those presented in this work is in the number of encapsulated dye molecules. This results in differences in FRET efficiency, fluorescence brightness, number of J and H dimers of the encapsulated dyes [31]. Fluorescence quenching of dimers and FRET are both temperature-dependent [45]. All these are conceivably the reasons for the observed enhanced sensitivity of UNSNTs. A substantial enhancement in the temperature resolution can be explained by the high fluorescence brightness of the synthesized UNSNTs. Because of their brightness, it was sufficient to use the excitation laser power that was approximately 80 times smaller than that in the study of nanothermometers of the previous record [38]. Since the temperature resolution is proportional to the power of the excitation light, the major improvement in the temperature resolution can be explained by their ultrahigh fluorescence brightness. The exact contribution of FRET, dimerization, quantum yield, and other parameters of the encapsulated dyes in the observed enhancement of temperature measurements will be investigated in future works.

## CONCLUSIONS

Here we presented the synthesis of ultrabright fluorescent nanothermometers (UNSNTs) that offer record sensitivity and thermal resolution. These 40 nm mesoporous silica particles encapsulate two fluorescent dyes: a temperature-dependent RB and a temperature-independent R6G. The temperature is measured as the ratio between fluorescence intensities in two spectral bands, achieving a relative sensitivity of 11±1.3 %K$^{-1}$, a 2-6x improvement over previous all-optical nanothermometers. The UNSNTs demonstrate a temperature resolution relative to power density of 128 K·Hz$^{-0.5}$·W·cm$^{-2}$, a 455x improvement compared to the thermometers reported in the literature. This record resolution enables extremely fast temperature monitoring (tens of microseconds to milliseconds). The improvements stem from the high density of fluorescent dye molecules within the nanoparticles, achieved by introducing small amounts of hydrophobic ethyl groups to the silica matrix, creating a Lotus effect that enhances pore hydrophobicity while maintaining a hydrophilic particle surface. This all-optical method is applicable to various transparent media and microscopy setups, with its ratiometric nature ensuring independence from excitation light power. The high sensitivity and speed make these nanothermometers promising for studying thermodynamics in complex biological structures, microfluidics, and various photonic and electronic devices, representing a significant advancement in rapid and accurate nanoscale temperature mapping crucial for various scientific and technological domains.

# Experimental section

**Chemicals and Materials**. The study employed the following reagents: Tetraethylorthosilicate (TEOS, purity ≥ 99%, acquired from Acros Organics, Fair Lawn, NJ, USA), Triethanolamine (TEA, reagent grade, 98% pure, sourced from Sigma Aldrich, St. Louis, MO, USA), Cetyltrimethylammonium bromide (CTAB, high purity grade, purchased from Amresco, Solon, OH, USA), Ethyltriethoxysilane (ETES, 96% pure, obtained from Frontier Scientific, Logan, UT, USA), Rhodamine 6G (R6G, provided by Sigma Aldrich), and Rhodamine B (RB, supplied by Exciton, Dayton, OH, USA). For dialysis, a RC membrane (Spectra/Pore, Rancho Dominguez, CA, USA) with a molecular weight cut-off range of 10–15 kDa was used. All synthesis processes utilized Deionized (DI) water.

**Nanothermometer Synthesis**: The creation of nanothermometers encapsulating R6G and RB followed a previously described procedure. Initially, CTAB (0.69 g, 1.9 mmol), R6G, and RB were dissolved in 21.7 mL of DI water at 90 °C. Simultaneously, a mixture of TEA (14.3 g, 96 mmol) and TEOS (1.71 g, 8.2 mmol) was heated at 90 °C for 3 hours. This mixture of TEOS and TEA was then combined with the solution containing the dyes and CTAB. After mixing for 30 minutes, ETES (196 µL, 0.9 mmol) was introduced, and the mixture was stirred for an additional 3 hours. A 1:1 molar ratio of R6G to RB was used.

The concentration of each dye was kept at 0.15 mmol. Excess reagents were removed using a dialysis membrane (with a molecular weight cut-off of 10–15 kDa). The dialysis was performed for 2-3 days by changing water every four hours (except overnight).

**Nanothermometer Characterization**: The absorbance spectra of the nanoparticles were measured using a Cary 60 UV-VIS spectrophotometer by Agilent, Inc. (Santa Clara, CA, USA). The spectral bandwidth and the scan rate of the UV-Vis measurements were 1.5 nm and 300 nm/min, respectively. Images of individual nanoparticles and fluorescence spectra were captured using a confocal Raman microscope by WITec, Inc. (now, a part of Oxford Instruments, UK). The integration time for taking images using confocal Raman microscope was 30 ms per spectrum. A Nikon objective was an oil-immersed 100X one with 1.3 NA. The nanoparticle size distribution was determined using dynamic light scattering (DLS) with a Zetasizer Nano ZS (Malvern Panalytical, UK). The concentration of nanoparticles suitable for DLS analysis was estimated by the absorbance of the particle dispersion in the range of 0.001-0.01 at 550 nm wavelength. An Icon atomic force microscope AFM (Bruker-Nano Inc, Santa Barbara, CA, USA) was used to image the particles. The AFM imaging was conducted in Peakforce QNM mode using Scanasyst-AIR probes. The imaging parameters were set to minimize the peak force error (may depend on a particular microscope and the environmental community). Fluorescence measurements of nanothermometer suspension were performed with a Cary Eclipse fluorescent spectrometer (Agilent, CA, USA). The slit width and scan rate of the Fluorescence measurements were 5 nm and 120 nm/min, respectively. Fluorescence ratiometric imaging was conducted using a laser scanning confocal microscope STELLARIS 8 (Leica Microsystems Inc., IL, USA). Specific accumulation times were mentioned in the text. Nanoparticles were deposited onto coverslip substrates (Fisherbrand by Fisher Scientific, Inc., USA) for the imaging. A Plan-Apochromat 40X Leica objective (N.A. 1.3) was utilized for the fluorescence ratiometric imaging. The laser power was measured by Thorlabs Power and Energy Meter (Thorlabs, Inc., USA), see the Supplementary information for detail. The calibration of the nanothermometers was done in water (or air) at different temperatures as measured by either a thermocouple (or a HTI-19 thermal camera by XT Instrument, Inc., China). It should be noted that the calibration of nanothermometers is valid for the medium with a relatively small change of the refractive index at the spectral ranges of R6G dye (510 nm - 560 nm) and RB dye (560 nm - 580 nm). It is quite a typical case for transparent blank media. Since the fluorescence intensity is proportional to the square of the refractive index, one can estimate a potential error in the calibration. For example, even in the case of glass prisms, in which one has a quite strong dependence of the refractive index on the wavelength of light (dispersion), the change of refractive index of the glass prisms (Koop Glass, Inc. Pittsburgh, PA) in the spectral range of the dyes is about 0.3%. It would change the fluorescence ratio by 0.6%. For the reported here nanothermometers, it is translated into the error of 0.06 K, which can be considered negligible.

**Acknowledgments**

Support of this work by NSF CBET grants 1911253 and 2110757 are acknowledged. The used confocal microscopes were purchased with the help of the NSF 1428919 grant and the Massachusetts Life Science Center grant, Bits-to-Bytes program.

**Conflict of Interest**

The authors declare no conflicts of interest.

**Data Availability Statement**

The data sets generated during and/or analyzed during the current study are available from the corresponding author upon reasonable request.

**Supplementary Information**

The Supplementary information is available free of charge at  https://....

AUTHOR INFORMATION

**Corresponding Author**

    **Igor Sokolov** – *Department of Mechanical Engineering and Biomedical Engineering and Physics, Tufts University, Medford, MA, USA.*
    Email: igor.sokolov@tufts.edu

**Authors**

    **Mahshid Iraniparast** – *Department of Mechanical Engineering, Tufts University, Medford, MA, USA.*
    **Nishant Kumar** – *Department of Mechanical Engineering, Tufts University, Medford, MA, USA.*

**Author Contributions**

I. S. conceived the idea of the project and supervised the project. M.I. fabricated, characterized the nanothermometers, and analyzed the data. N.K. did the AFM characterization. M.I. and I.S. wrote the manuscript. All authors discussed the results and commented on the manuscript.

# Supporting information

# for

# Single Ultrabright Fluorescent Silica Nanoparticles Can Be Used as Individual Fast Real-Time Nanothermometers

Mahshid Iraniparast, [1], Nishant Kumar, [1] Igor Sokolov [1,2,3,*]

[1] Department of Mechanical Engineering, Tufts University, Medford, MA 02155, USA;

[2] Department of Biomedical Engineering, Tufts University, Medford, MA 02155, USA.

[3] Department of Physics, Tufts University, Tufts University, Medford, MA 02155, USA.

## Calculating the number of dye molecules within each nanoparticle

The Beer–Lambert law was employed to calculate the number of dye molecules per volume in this study. The UV-Vis absorbance of nanoparticles encapsulated with R6G and RB was measured for this purpose. The extinction coefficients of R6G and RB, as referenced in the previous study (ref. 33 the manuscript), were used in determining the concentration of each of the dye molecules. The result showed $3.91 \times 10^{13}$ of R6G and $4.71 \times 10^{13}$ of RB per volume.

Subsequently, the total number of particles per volume was calculated based on the density, weight, and volume of nanoparticles. The resulting total number of particles was determined to be $3.087 \times 10^{10}$ per volume. Dividing the total number of dye molecules per volume by the total number of particles per volume yielded the number of dye molecules per particle. The detailed results are presented in the main content of the study. The whole calculation is described below:

$$Beer\ Law: A = \varepsilon \times C \times L$$

$$0.005865 = 90300 \times C \times 1$$

$$C = 6.49 \times 10^{-8}\ (Mol)$$

$$Number\ of\ R6G\ dye\ molecules\ per\ volume: 6.49 \times 10^{-8} \times 6.022 \times 10^{23} \times 10^{-3}$$
$$= 3.91 \times 10^{13}$$

$$Weight\ of\ one\ particle: \rho \times \frac{4}{3} \times \pi \times R^3 = 1600 \times \frac{4}{3} \times \pi \times (\frac{43}{2})^3 \times 10^{-6}$$
$$= 6.663 \times 10^{-14}\ (mg)$$

$$Number\ of\ particles\ per\ volume: \frac{\frac{(12 \times 0.001 \times 0.6)}{3.5}}{6.663 \times 10^{-14}} = 30872067297$$

$$Number\ of\ R6G\ per\ each\ nanoparticle: \frac{3.9119 \times 10^{13}}{30872067297} = 1267.1482$$

The same approach was used for calculation of number of RB molecules within each nanoparticle:

$$Beer\ Law: A = \varepsilon \times C \times L$$

$$0.00735 = 94000 \times C \times 1$$

$$C = 7.819 \times 10^{-8}\ (Mol)$$

$$Number\ of\ RB\ dye\ molecules\ per\ volume: 7.819 \times 10^{-8} \times 6.022 \times 10^{23} \times 10^{-3}$$
$$= 4.7094 \times 10^{13}$$

$$Weight\ of\ one\ particle: \rho \times \frac{4}{3} \times \pi \times R^3 = 1600 \times \frac{4}{3} \times \pi \times (\frac{43}{2})^3 \times 10^{-6}$$
$$= 6.663 \times 10^{-14}\ (mg)$$

$$Number\ of\ particles\ per\ volume: \frac{\frac{(12 \times 0.001 \times 0.6)}{3.5}}{6.663 \times 10^{-14}} = 30872067297$$

$$Number\ of\ RB\ per\ each\ nanoparticle: \frac{4.7094 \times 10^{13}}{30872067297} = 1525.4804$$

## Calculating the fluorescence brightness of the nanothermometers

The brightness of individual nanoparticles was determined by calculating their relative brightness compared to standard fluorophores. In this case, the standard fluorophores are naturally free rhodamine 6G and B dyes dissolved in water. It well describes the brightness of the particles at different wavelengths because the fluorescent spectra of individual dye molecules are almost unchanged after encapsulation. Specifically, the fluorescence spectra of rhodamine 6G after encapsulation have a small redshift of about 5 nm, whereas rhodamine B demonstrates a blueshift of 3 nm. [1, 2] The used unit of measurement for brightness is MESF (Molecules of Equivalent Soluble Fluorochrome), and it is obtained using the following equation:

$$Relative\ Brightness = \frac{FL_{NPs}/Number\ of\ nanoparticles}{FL_{dye}/Number\ of\ dye\ molecules}$$

Here, $FL_{NPs}$ and $FL_{dye}$ represents the fluorescence intensity of nanoparticles and dye molecules, respectively. The brightness was calculated separately for each of the two molecules, R6G and RB, and the results are mentioned in the main content.

## Measurements of laser power density used for the imaging in scanning laser confocal microscope

The laser power produced by the supercontinuum laser of the Leica laser scanning confocal microscope was measured using a Thornlab power meter tuned for the excitation wavelengths of 488 nm. The detection head was placed right above the microscope (inverse) 40x, 1.3 NA objective. The distance between the head and the objective was chosen to maximize the power reading (and it was constant with at least ~2mm interval of the distances). The power was measured at several positions of the laser power control: 2, 3, 4, 8% (the percentage of the total working power of the microscope laser). All measurements presented in the paper were down at the 2% of the laser power (with the exception of the measurements of the dependence of the fluorescence ratio on the laser power). The results are presented in Table S1.

**Table S1**. The results of the measurements of the laser power and laser power density of the super continuum laser used in the Leica laser scanning confocal microscope.

| Laser power control % | Measured power [µW] | Power density W/cm² |
|---|---|---|
| 2 | 0.7 | 425 |
| 3 | 1.1 | 638 |
| 4 | 1.4 | 850 |
| 8 | 2.8 | 1700 |

The power density was calculated as the measured power divided by the size of the laser spot, which is the airy disc = 1.22 $\lambda$/NA, where $\lambda$=488 nm, NA=1.3. Note that the same method was used in the paper which reported the previous record [reference 18 of the main text].

## Calculation of the temperature resolution

An example of fitting of temperature fluctuations $\sigma_T$ versus the time of measurement $t_m$ is shown in figure S1a. The measured temperature resolutions from individual nanoparticles are presented in Figure S1b. It is important to stress that this quite noticeable spread of the temperature resolutions (Figure S1b) does not have a direct relation to the accuracy of the conversion of the fluorescence ratio to temperature, which can be presented by the relative sensitivity. This is because the spread of temperature resolutions is defined by the total number of fluorescent molecules in each particle, or the volume of the particle. For example, a large particle could be considered as several smaller particles together. As one can see from Figure 3c, the larger number of particles, the smaller the temperature uncertainty (simply due to the ergodic averaging), and consequently, the higher temperature resolution (see, also Fig.4a). If the dye molecules are spread evenly through the particle volume, the accuracy of finding temperature using the described ratiometric approach is independent of the particle volume. Indirectly, it manifests itself by the narrowness of the relative sensitivity, Figure 3b.

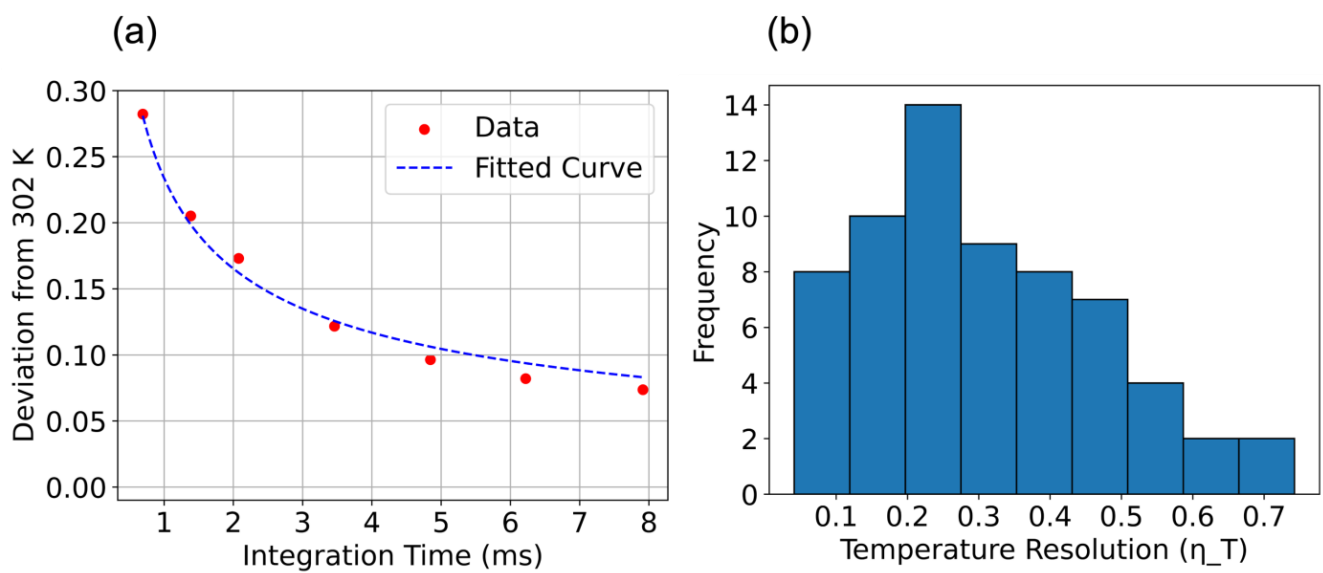

**Figure S1. Calculation of the temperature resolution $\eta_T$:** (a) an example of fitting of temperature fluctuations $\sigma_T$ versus the time of measurement $t_m$. (b) The histogram of the temperature resolutions for individual nanothermometers. Sixty individual nanoparticles were investigated. The mode and average are 0.25 K.Hz$^{-1/2}$ and 0.30 K.Hz$^{-1/2}$, respectively. All the measurements were performed using excitation power density of $455 (w.cm^{-2})$.

## Calculation of fluorescence ratio – temperature dependence

The measured data is shown in figure S2. The temperature dependence was calculated using the mean least square method using the measurement data shown in Figure S2. Formula $Temperature = 10.3 \times Ratio + 293$ was derived for the linear regression of the mean least square method.

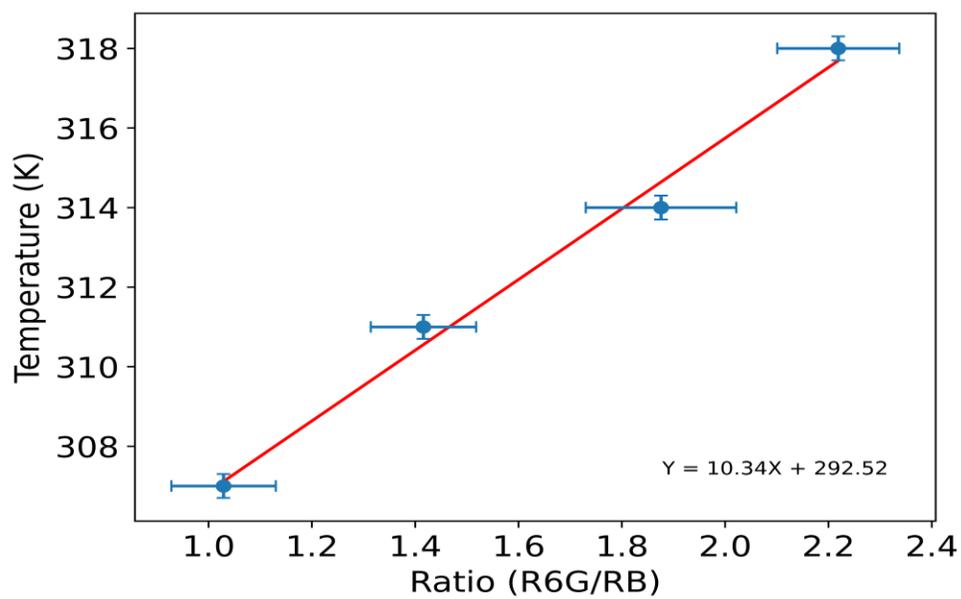

**Figure S2.** The fluorescence ratio for the individual nanothermometers upon changing temperature. The plot demonstrates a linear relationship between the change in the R6G to RB fluorescence intensity ratio and temperature. Each data point is calculated for ten different nanothermometers; the fluorescence signal was collected for 20 ms for each nanothermometer. This data was collected using the laser scanning confocal microscope (Leica).

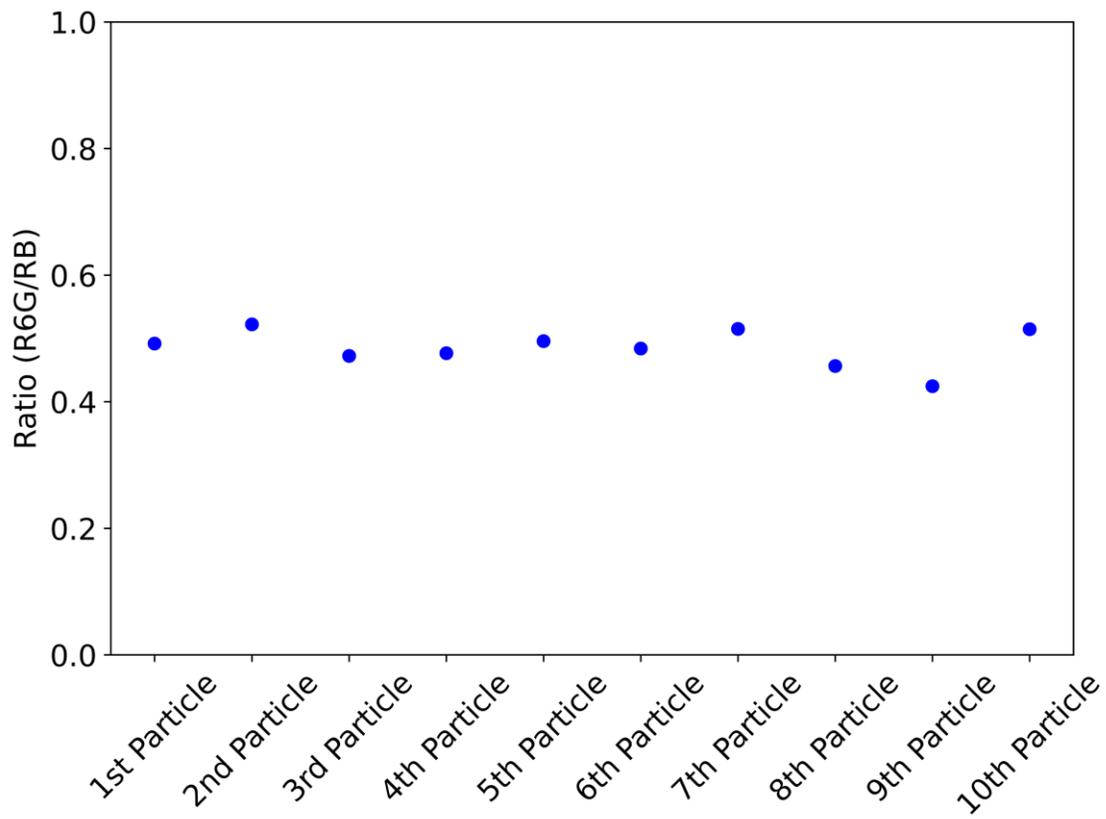

**Figure S3:** Fluorescence intensity of R6G to RB ratio of 10 random individual nanothermometers. This graph is prepared using the data shown in figure 3b. The collection time for each nano thermometer was 30 ms. These data were collected using the Raman confocal microscope.

## Photoluminescence stability of the nanothermometers

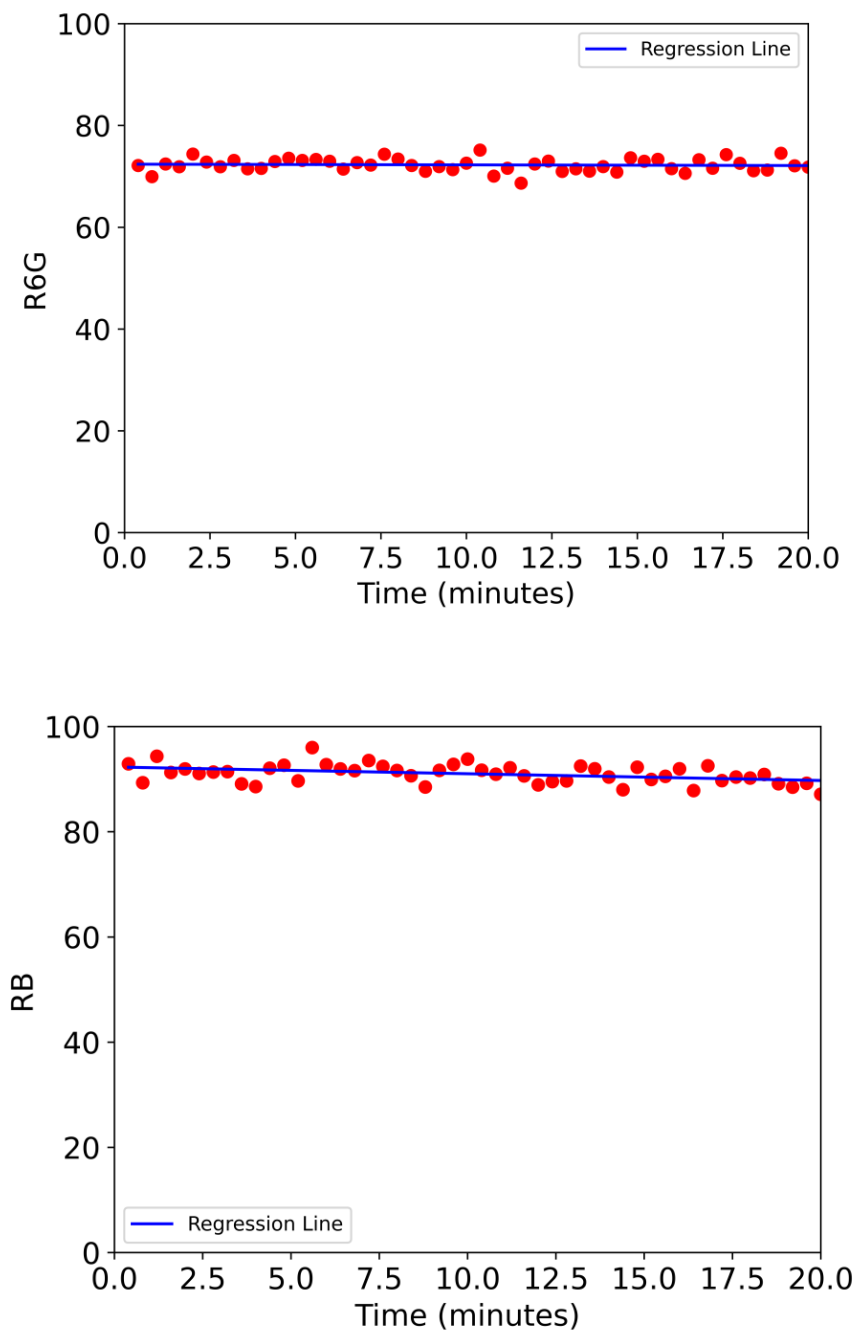

**Figure S4: Photoluminescence stability:** Fluorescence intensity of the measured nanothermometers integrated in the spectral range of R6G dye (510 nm - 560 nm) and RB dye (560 nm - 580 nm).

# Demonstration of the presence of the Förster resonance energy transfer (FRET) between the dyes encapsulated inside of the nanothermometers

R6G and RB dyes were dissolved in water in the same proportion as encapsulated inside the nanothermometers. The emission–correlation matrices are shown in supplementary Figure S5 for both water-diluted dyes and the same proportion of the dyes encapsulated in the nanothermometers. To ensure that there was no FRET in the water-dissolved dyes, the concentration of the dyes was ~ $10^3$ lower compared to the dyes encapsulated in the particles (of the order of micromoles). One can see from supplementary figure S5 that the fluorescence emitted by R6G molecules is substantially depleted in the case of the dyes encapsulated inside nanoparticles, whereas the fluorescence intensity of RB dye is substantially increased. This is what should be expected as a result of FRET: R6G serves as a donor, which transfers its energy to the acceptor, RB dye.

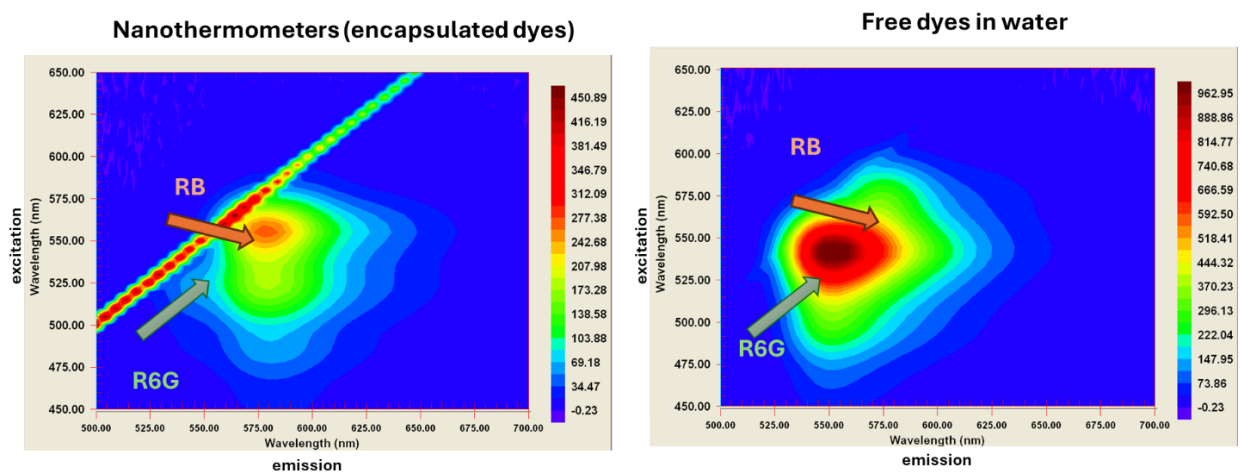

Figure S5: The emission-excitation matrix of the mix of the two dyes encapsulated inside the nanothermometers and dissolved in water. Dyes were dissolved in water at the same proportion as encapsulated inside the nanothermometers. The arrow shows the directions towards the location of RB and R6G peaks. (The absence of clear locations of individual peaks are explained by the spectral overlap between the two dyes.)

Another reason for FRET can be given by estimating the distances between the dye molecules inside of each particle. To calculate it, we assume a homogeneous distribution of the dye molecules across the silica matrix of the nanoparticle (which is a reasonable assumption; see ref. 31 the manuscipt). Taking the average diameter of the particle of 43 nm and the total number of molecules per particle to be 2800 (1270 of R6G and 1530 of RB molecules), one can obtain the average distance between the molecules of 5 nm. On the other hand, one can estimate the distance between molecules assuming that they are located in the cylindrical silica channels inside the particle. It is characterized by the DFT pore diameter of 3.8 nm, and the available pore volume of 0.75 cm³/g (ref. 35 the manuscript). Assuming a homogeneous distribution of the same number of dye molecules along the cylindrical channels, one can obtain an average distance between the dye molecules of 1.6 nm. If we assume for simplicity that each pair of these molecules is donor and acceptor, we can obtain the FRET efficiency using the equation:

$$\text{Efficiency} = R_0^6/(R_0^6 + r^6), \quad (S1)$$

where R_0 is the Förster distance and r is the distance between dye molecules. R0 was calculated to be 8.79 nm (see, e.g., ref. 32 of the manuscript), considering the emission spectrum of R6G dye as a donor and the absorbance spectrum of RB as an acceptor. Using this equation, one can estimate the FRET efficiency as 97-100%.

However, the above estimation is not entirely correct. It can only be referred to as the ideal case of close proximity of donor and acceptor molecules. When speaking about FRET efficiency between molecules encapsulated in a nanoparticle, one has to take into account the random location of the molecules, which results in a limited number of donor – acceptor pairs. As was shown as a result of statistical simulations in [1] (ref. 33 the manuscript), the effective FRET between R6G and RB dyes encapsulated inside of a nanoparticle is substantially smaller than the estimated using the idealized case of donor-acceptor pairs. Running the simulations described in [1] for the number of dye molecules measured in this work, one can find the efficiency FRET has ~41% efficiency.

Finally, FRET can also be estimated using the following formula:

$$\text{Efficiency} = 100\% * (1 - \frac{FI_{in\,presence\,of\,acceptor}}{FI_{no\,acceptor}}), \quad (S2)$$

where $FI_{in\ presence\ of\ acceptor}$ is the fluorescence intensity in the presence of the acceptor, and $FI_{no\ acceptor}$ is that in the absence of the acceptor. The mesoporous silica nanoparticles particles containing only donor (R6G) in the appropriate concentrations range were investigated in [2] (ref. 31 of the manuscript). Although such high concentrations were not reached in [2], one can see a linear dependence between the dye concentration and brightness of mesoporous silica nanoparticles containing R6G (Fig.2 of ref. [2]), we can estimate the brightness of such particles containing 1270 molecules of R6G as 1050 (Brightness relative to one molecule of R6G). Therefore, one can estimate $FI_{no\ acceptor} = 1050$. The brightness of the particles reported here (which contain 1270 molecules of R6G in presence of RB acceptor) is equal to 650 in the same units. It gives the efficiency of FRET of 36%, which is in relatively good agreement with the results of the simulations given above.